
\input phyzzx
\font\tenrm=cmr10
\overfullrule=0pt
\pubnum{\vbox{\hbox{TIFR/TH/93-36}\hbox{hep-th/9308062}}}
\date{August 1993}
\titlepage
\title{PERTURBATION OF THE GROUND VARIETIES
OF {\it C}~=~1 STRING THEORY}

\author{Debashis Ghoshal, Porus Lakdawala and Sunil Mukhi}
\foot{e-mail: (ghoshal/porus/mukhi)@theory.tifr.res.in}
\vskip 5pt
\address{\vbox{\hbox{Tata Institute of Fundamental Research}
\hbox{Homi Bhabha Road, Bombay 400005, India}}}
\vskip 8pt
\vskip 2truecm
\abstract{
We discuss the effect of perturbations on the ground rings of $c=1$
string theory at the various compactification radii defining the $A_N$
points of the moduli space. We argue that perturbations by plus-type
moduli define ground varieties which are equivalent to the unperturbed
ones under redefinitions of the coordinates and hence cannot smoothen
the singularity. Perturbations by the minus-type moduli, on the other
hand, lead to semi-universal deformations of the singular varieties
that can smoothen the singularity under certain conditions. To first
order, the cosmological perturbation by itself can remove the
singularity only at the self-dual ($A_1$) point.}
\vfill\eject

\def\co{{\cal O}}
\def\bco{\overline{\cal O}}
\def\nby2{{N\over 2}}
\def\Z{{\bf Z}}
\def\C{{\bf C}}
\def\I{{\cal I}}
\def\G{\Gamma}
\def\D{\Delta}
\def\wbar{\overline {W}}
\def\ybar{\overline {Y}}
\def\del{\partial}
\def\tet{{\cal T}}
\def\ha{{1\over2}}
\def\np{Nucl. Phys.}
\def\pl{Phys. Lett.}

\def\bb{{\bar b}}

\subsection{\bf Introduction}

The study of polynomial rings associated to systems with a
BRS cohomology is a useful way to understand their physical
properties\REF\chiralrings{W. Lerche, C. Vafa and N. Warner,
Nucl. Phys. {\bf B324} (1989) 427.}\REF\witten{E. Witten, Nucl. Phys.
{\bf B373} (1992) 187.}[\chiralrings,\witten]. For noncritical $c=
1$ string theory, the ring associated to cohomology states of ghost
number zero[\REF\lzetal{B. Lian and G. Zuckerman, Phys. Lett. {\bf B266}
(1991) 21,\hfill\break S. Mukherji, S. Mukhi and A. Sen,
Phys. Lett. {\bf B266} (1991) 337,\hfill\break
P. Bouwknegt, J. McCarthy and K. Pilch,
Comm. Math. Phys. {\bf 145} (1992) 541.}\lzetal]
is particularly interesting, as it gives an insight into
the structure of the unbroken gauge symmetries of this
backgrounds\REF\witzwie{E. Witten and B. Zwiebach, \np\ {\bf 377}
(1992) 55.}[\witten,\witzwie]. It turns out that the polynomial ground
ring defines a singular variety, and the unbroken symmetries have a
natural action on this variety as volume-preserving diffeomorphisms.

The $c=1$ string has a large collection of marginal deformations generated
by its various moduli. The best known of these moduli are the cosmological
operator, the radius-changing operator, and the operator which deforms the
background into a black hole\REF\blackholepert{G. Mandal, A.  Sengupta
and S. Wadia, Mod.  Phys. Lett. {\bf A6} (1991)
1685.}\REF\mmsbh{S. Mukherji, S. Mukhi and A. Sen,
Phys. Lett. {\bf B275} (1992)
39.}[\blackholepert,\mmsbh].
The effects of the cosmological perturbation at the
self-dual point and in the uncompactified theory have been examined in%
\REF\kms{D. Kutasov, E. Martinec and N. Seiberg, \pl\ {\bf B276}
(1992) 437.}\REF\kachru{S. Kachru, Mod. Phys. Lett. {\bf A7} (1992)
1419.}\REF\barbon{J. Barb\'on, Int. J. Mod. Phys. {\bf A7} (1992)
7579.}\ Refs.[\witten,\kms--\barbon]. The moduli space of the
$c=1$ string generated by the radius-changing operator, and the nature
of the polynomial ground ring and its associated variety at various
points of this moduli space, have been analyzed in[\REF\gjm{D.
Ghoshal, D.P. Jatkar and S. Mukhi, Nucl. Phys.  {\bf B395} (1993)
144.}\gjm] and found to be related to some beautiful mathematical
structures, the Kleinian singular varieties. One can explicitly see
how the symmetries vary as a function of the compactification radius.
An equally explicit understanding of the symmetry-breaking pattern
along general directions in the moduli space is, however, lacking so
far.

In this note we uncover some aspects of the nature of ground rings and
their associated varieties when the $c=1$ string is perturbed by
various generic moduli. For this purpose, we start with the theory
defined at some integer multiple of the self-dual radius,
corresponding to the $A_N$ points. Possible perturbations fall into
two classes, generated by the plus- and minus-type moduli.  We will
argue that generic minus-type perturbations smoothen the singularities
of the ground varieties, but plus-type perturbations cannot do so.
Also, the cosmological perturbation alone does not effect a
smoothening of the singularity. We will also find an intriguing
relation between the minus-type perturbations and the theory of
semi-universal deformations of Kleinian singularities.

Let us now briefly recall some facts about $c=1$ string theory. The
matter sector is described by the CFT of a compact free boson $X$.
The moduli space of $c=1$ CFT is well known%
\REF\ginsparg{P. Ginsparg, Nucl. Phys. {\bf B295} (1988) 153,\hfill\break
G. Harris, Nucl. Phys. {\bf B300} (1988) 588.}[\ginsparg]. At the
self-dual radius, the theory has an enhanced $SU(2)\otimes SU(2)$
symmetry, and the (chiral) operators are labelled by their $SU(2)$
quantum numbers $s=0,\half,1,\cdots$ and $-s\le n\le s$: $V_{s,n}$.
When this CFT is coupled to gravity, the operators $V_{s,n}$ are
``dressed'' by Liouville vertex operators which can have two possible
momenta $p^\varphi_\pm = i\sqrt 2(-1\pm s)$. The former is called the
{\it plus-type} dressing and the latter {\it minus-type}.

The BRS analysis[\lzetal] shows that apart from the operators
$Y_{s,n}^\pm = c V_{s,n} e^{\sqrt 2(1\mp s)\varphi}$ of standard ghost
number 1 (which exist for either dressing) there exist an infinite
number of operators $\co^{(+)}_{s,n}$ and ${\cal P}^{(-)}_{s,n}$ at ghost
numbers 0 and 2 respectively. The operators relevant for the closed
string theory are constructed by combining the chiral and anti-chiral
operators. Since the Liouville field is non-compact,
its left and right momenta must be matched. This results in
plus-type operators of ghost numbers 0, 1 and 2 and minus-type of
ghost numbers 2, 3 and 4.

There also exist additional operators (``new moduli'') which are in
the relative cohomology of $b_0 - \bb_0$ but not of $b_0$ and $\bb_0$
separately\REF\miaoli{M. Li, Nucl. Phys. {\bf B382} (1992)
242.}[\miaoli,\witzwie]. These can be expressed as BRS commutators of
terms explicitly involving the Liouville field $\phi$, hence it has
been argued that they are not genuinely BRS-exact. However, it has
been pointed out\REF\maha{S. Mahapatra, S. Mukherji and A. Sengupta,
Mod. Phys. Lett. {\bf A7} (1992) 3119.}[\maha] that in string field
theory such operators are normally taken to be BRS-trivial, as they do
not generate new deformations of the theory. In what follows, we will
therefore work with the cohomology in which these new moduli are set
to zero.

The ghost number 0 operators $\co_{s,n}\overline{\co}_{s,n'}$ form a
ring, called the ground ring,
with the OPE (modulo BRS exact terms) defining the ring
multiplication. The ring at the $SU(2)$ point is generated by the four
operators (two electric and two magnetic)
$$
a_1 = \co_{\ha,\ha}\overline{\co}_{\ha,\ha}\qquad
a_2 = \co_{\ha,-\ha}\overline{\co}_{\ha,-\ha}\qquad
a_3 = \co_{\ha,\ha}\overline{\co}_{\ha,-\ha}\qquad
a_4 = \co_{\ha,-\ha}\overline{\co}_{\ha,\ha}
\eqn\oobar
$$
satisfying the relation
$$
a_1 a_2 - a_3 a_4 = 0.
\eqn\gcone
$$
which define a singular conical variety, the ``ground cone''.

The ghost number 1 operators $J_{s+1,n,n'} \equiv
Y^+_{s+1,n}\overline{\co}_{s,n'}$ and $\bar J_{s+1,n,n'} \equiv
\co_{s,n}\overline {Y}{}^+_{s+1,n'}$ act on the ground ring as
generators of volume-preserving diffeomorphisms of the ground cone.
Finally, the ghost number 2 operators $Y_{s,n}^+\overline {Y}{}^+_{s,n'}$
and $Y_{s,n}^-\overline {Y}{}^-_{s,n'}$ correspond to the moduli since
they give rise, via the descent equation, to ghost number 0 two-forms
$W^\pm_{s,n,n'}dz\wedge d\bar z\equiv V_{s,n}V_{s,n'}
                   e^{\sqrt 2(1\mp s)\varphi} dz\wedge d\bar z$,
that correspond to marginal deformations of the action.

All of the above applies to the theory at the self-dual radius.
Consider now the special set of $A_N$ points correponding to
compactification radius $N/\sqrt 2$, which arise on modding out the
self-dual theory by the discrete subgroups $\Gamma=\Z_{2N}$ of $SU(2)$%
[\ginsparg].
The operators at these points satisfy a constraint arising from
the matching of the left and right matter momenta
$$
n-n' = 0 \quad{\rm mod}~~N.
\eqn\matter
$$
It was shown[\gjm]
that the generators of the ground ring at each of these points are the
$\Z_{2N}$-invariant polynomials in the generators \oobar\ at the $SU(2)$
point, namely,
$$
a_1 = \co_{\ha,\ha}\overline{\co}_{\ha,\ha}\qquad
a_2 = \co_{\ha,-\ha}\overline{\co}_{\ha,-\ha}\qquad
X = \co_{\nby2,\nby2}\overline{\co}_{\nby2,-\nby2}\qquad
Y = \co_{\nby2,-\nby2}\overline{\co}_{\nby2,\nby2}
\eqn\generators
$$
satisfying the relation
$$
(a_1 a_2)^N - XY = 0.
\eqn\gvar
$$
which defines the $A_N$ ground variety. While the ground cone \gcone\
has an isolated singularity at the origin, the singular locus of the
$A_N$ variety \gvar\ (except for $N$=1) is given by the pair of
straight lines defined by $a_1a_2=0$.

While eq.\matter\ forbids some of the moduli in the orbifold theory,
one gets additional moduli in the twisted sector
$c\bar c\tet^\pm_q =
c\bar c e^{i\sqrt 2q X + \sqrt 2(1\mp q)\varphi}$,
$q\in \Z^+ + {1\over N},{2\over N},\cdots,{N-1\over N}$.
We call these the intermediate tachyons. In the limit of infinite
radius ($N\to\infty$), the intermediate
tachyons form a continuum of operators labelled by a real number $q$.
There are, however, no states with non-standard ghost number in the
twisted sector.

It was argued[\kms] that the ghost number 1 operators and the
plus-type moduli, including the intermediate tachyons, form a module
for the ground ring (modulo BRS exact terms) under the OPE. More
precisely, the plus-type discrete moduli form a faithful module and
the plus-type intermediate tachyons form unfaithful modules for each
fractional part of $q$.

\subsection{\bf Deformation Of The Ground Varieties}

In this section we will study the effect of perturbations by various
marginal operators on the ground rings and associated varieties of $c=1$
string theory. Our arguments are based on momentum
counting, and we will restrict ourselves to the $A$-type points in the
moduli space of the $c=1$ CFT.

We will use the Hilbert space of the unperturbed theory to describe
the effects of the perturbations, under the assumption that the theory
changes smoothly. This needs some justification as we know that at
least for the radius perturbation the cohomology varies in a singular
fashion[\gjm]. For the perturbations that we are interested in, our
assumption can be made plausible by appealing to their connection to
the deformation of the corresponding Kleinian singularities.

Let us begin with the simplest case: the cosmological perturbation at the
$SU(2)$ point, induced by adding the operator $\mu\int W_{0,0,0}$ to the
action. (This is the unique operator for which the plus and minus dressings
are equivalent.) In Ref.[\witten], it was argued that under this
perturbation, the $SU(2)$ ground cone
\gcone\ is deformed to the smooth variety defined by
$$
a_1 a_2 - a_3 a_4 = \mu.
\eqn\pcone
$$
This observation was motivated by an analysis of the $SU(2)\otimes SU(2)$
content of the moduli. Let us give an alternative motivation for the same
result, using just the matching of Liouville and matter momentum on both
sides of the ground ring relation. The Liouville and matter momenta of the
cosmological operator are given by $(p^\varphi,p_L^X, p_R^X) = \sqrt2(-i,
0, 0)$. If we associate the negative of this momentum to the parameter
$\mu$ then the perturbation formally conserves momentum.  Now, the momenta
carried by $a_1 a_2$ (and $a_3 a_4$) are $(p^\varphi,p_L^X, p_R^X) =
\sqrt2(i, 0, 0)$, precisely the same as those we have identified with the
parameter $\mu$, hence the perturbed equation \pcone\ is the unique one
consistent with momentum conservation. Of course, it is assumed that the
coefficient of the allowed perturbation to the ring relation does not
accidentally vanish, since there is no known reason for it to do so.
(The above argument of matching of Liouville momenta is
just a scaling argument in the spirit of Refs.[\REF\kpz{V.
Knizhnik, A.M. Polyakov and A. Zamolodchikov, Mod. Phys. Lett. {\bf
A3} (1988) 819.}\kpz,\REF\ddk{F. David, Mod. Phys. Lett. {\bf A3}
(1988) 1651,\hfill\break J. Distler and H. Kawai, Nucl. Phys. {\bf
B321} (1989) 509.}\ddk].)

Remaining at the $SU(2)$ point, we can now consider more general
perturbations. Suppose first that we perturb by a general modulus of
plus-type:
$$
S\to S + \sum_{s,n,n'} u^+_{s,n,n'} \int W^+_{s,n,n'}.
\eqn\pluspert
$$
The momenta associated with the parameter $u^+_{s,n,n'}$ are
$\sqrt2(i(1-s), -n, -n')$. Thus, each occurrence of this parameter in the
ring relation must be compensated by appropriate powers of the $a_i$.
Restricting for simplicity to the case $n=n'$, we find that the
equation of the perturbed ground cone consistent with momentum
conservation is
$$
 a_1 a_2 - a_3 a_4 = \sum_{s,n} \alpha_{s,n}
u^+_{s,n,n} a_1^{s+n}a_2^{s-n}.
\eqn\pconeplus
$$
The $\alpha_{s,n}$ are some constants, which are undetermined by
the momentum-matching considerations. Assuming again that they are
nonzero, they can be absorbed into the definition of the parameters $u^+$.
(The restriction of this equation to $n=0$ was also conjectured in[\witten].
There it was argued that among the perturbations by plus-moduli, those
with $n=n'=0$ are the only ones consistent with integrability, since they
commute among themselves --- the chiral part $W^+_{s,n}$ of these operators
lie in the Cartan subalgebra of $w_\infty$.)

Next we consider perturbing the $SU(2)$ theory by minus-moduli:
$$
S\to S + \sum_{s,n,n'} u^-_{s,n,n'} \int W^-_{s,n,n'}.
\eqn\minuspert
$$
The momenta associated to the parameters $u^-_{s,n,n'}$ are
$\sqrt2(i(1+s), -n, -n')$, as a result of which the analog of
eq.\pconeplus\ above would be obtained by the replacement $s\to -s$. This
would mean, however, that the ring generators $a_1$ and $a_2$ would be
raised to {\it negative} powers. Let us at this point make the
plausible assumption that the perturbed ground ring is also polynomial.
In this case the only allowed term is $s=n=0$, which is precisely the
cosmological perturbation. We conclude, therefore, that except for the
cosmological operator, the minus-type moduli cannot perturb the ground
cone at the $SU(2)$ point.

Things are quite different at the other $A_N$ points. Repeating the same
analysis, we easily find that the plus-type moduli can generate
perturbations of the form
$$
 (a_1 a_2)^N - XY = \sum_{s,n}
u^+_{s,n,n} a_1^{N+s+n-1}a_2^{N+s-n-1},
\eqn\pvarplus
$$
while a {\it finite number} of minus-type moduli can also generate
perturbations, of the form
$$
 (a_1 a_2)^N - XY = \sum_{0\le s-n, s+n\le N-1}
u^-_{s,n,n} a_1^{N-s+n-1}a_2^{N-s-n-1}.
\eqn\pvarminus
$$
Thus the set of minus moduli which can perturb the $A_N$ ground variety
fall into the ``diamond'' pattern displayed in Fig.1. There are precisely
$N^2$ minus-type perturbations at the $A_N$ point.

\midinsert
$$
\matrix{
{n\hfill\atop n'}&{}&{}&{}&{}&{}&\circ&{}&{\cdots}&{}&\circ&{}&{\cdots}
&{}&\circ&{}&{}\cr
\uparrow\hfill&{}&{}&{}&{}&\nearrow&{}&\searrow&{}&\nearrow&{}&\searrow
&{}&\nearrow&{}&{}&{}\cr
{}&{}&{}&{}&\bullet&{}&{}&{}&\circ&{}&{}&{}&\circ&{}&{\cdots}&{}&{}\cr
{}&{}&{}&\nearrow&{}&\searrow&{}&\nearrow&{}&\searrow&{}&\nearrow&{}
&\searrow&{}&{}&{}\cr
{}&{}&\bullet&{}&{}&{}&\bullet&{}&{}&{}&\circ&{}&{}
&{}&\circ&{}&{}\cr
{}&\nearrow&{}&\searrow&{}&\nearrow&{}&\searrow&{}&\nearrow&{}&\searrow&{}
&\nearrow&{}&{}&{}\cr
\bullet&{}&{}&{}&\bullet&{}&{}&{}&\bullet&{}&{}&{}&\circ&{}&{\cdots}
&\longrightarrow&s\cr
{}&\searrow&{}&\nearrow&{}&\searrow&{}&\nearrow&{}&\searrow&{}&\nearrow&{}
&\searrow&{}&{}&{}\cr
{}&{}&\bullet&{}&{}&{}&\bullet&{}&{}&{}&\circ&{}&{}
&{}&\circ&{}&{}\cr
{}&{}&{}&\searrow&{}&\nearrow&{}&\searrow&{}&\nearrow&{}&\searrow&{}
&\nearrow&{}&{}&{}\cr
{}&{}&{}&{}&\bullet&{}&{}&{}&\circ&{}&{}&{}&\circ&{}&{\cdots}&{}&{}\cr
{}&{}&{}&{}&{}&\searrow&{}&\nearrow&{}&\searrow&{}&\nearrow&{}
&\searrow&{}&{}&{}\cr
{}&{}&{}&{}&{}&{}&\circ&{}&{\cdots}&{}&\circ&{}&{\cdots}&{}&\circ&{}&{}\cr
}
$$
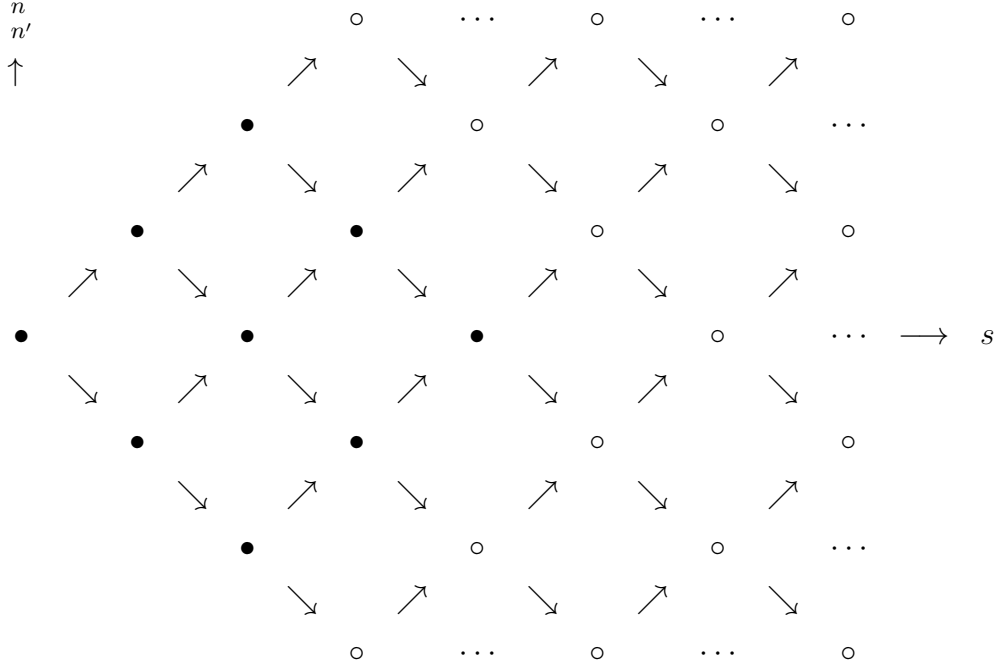
\centerline{Fig.1: The ``diamond'' at the $A_3$ point}
\centerline{\tenrm(The perturbations corresponding to the semi-universal
deformation are denoted by $\bullet$ and the rest by $\circ$)}
\endinsert

We notice here that in the case of minus-type perturbations, the
restriction to the electric moduli with $n=n'$, involves no loss of
generality. This is because of the fact that at the $A_N$ point, we have
the condition \matter.
Momentum counting leaves us with operators in the ``diamond'', that is
operators that satisfy the conditions $0\le (s-n),(s+n)\le N-1$. It is
easy to see that there are no magnetic operators in the ``diamond''
satisfying Eq.\matter.

Finally, one may consider perturbing the theory by the intermediate
tachyons $T^\pm_q$. In this case, momentum counting shows
that the perturbation to the ground variety can only be by fractional
powers of the $a_i$. We conclude that in first order, the intermediate
tachyon perturbations do not affect the ground variety.

\subsection{\bf Use of the $w_\infty$ symmetries}

We now show how the symmetries $J_{s,n,n'}=W^+_{s,n}\bco_{s,n'}$ act on
the moduli $W^-_{s,n}\wbar{}^-_{s,n'}$ and relate the different moduli of
the theory.  In particular, all the operators in the ``diamond'' with
non-zero matter momenta are related to those with zero matter momenta.
Furthermore, starting from the negative moduli in the ``diamond'' with
zero matter momenta, one generates all the moduli in the ``diamond'' and
no others. This symmetry can be used to calculate the effect of
perturbation by moduli with non-zero matter momenta, from the knowledge of
those with zero matter momenta.

Recall that the OPE between the chiral operators $Y^+=cW^+$ and $Y^-=cW^-$
gives rise to the following commutator[\witten]
$$
[W^+_{s_1,n_1},W^-_{s_2,n_2}] =
\cases{(s_1n_2+(s_2+1)n_1)\;W^-_{s_2-s_1+1,n_1+n_2},&if $s_2>s_1-1$ and\cr
{}&$|n_1+n_2|\le s_2-s_1+1$;\cr 0&otherwise.\cr}
\eqn\symm
$$
Since at the $A_N$ point, the modulus $W^-_{N-1,0,0}$ at the apex of the
``diamond'' has the maximum spin $s_2=N-1$, we can restrict our
considerations to symmetries with spin $s_1\le N-\half$. Let us for
definiteness, illustrate this with the example of the $A_2$ point. Here the
zero momentum operators in the ``diamond'' are the cosmological operator
$W^-_{0,0,0}$ and the operator $W^-_{1,0,0}$ that correspond to the
black-hole perturbation[\blackholepert,\mmsbh].
Since the maximum value of $s_2$
is 1, the relevant symmetries for this theory are $J_{{3\over
2},\pm{1\over2},\pm{1\over2}}$ and $J_{{3\over 2},\pm{3\over
2},\mp{1\over2}}$. The second set of operators annihilate $W^-_{1,0,0}$,
while the first set produce the states
$W^-_{{1\over2},\pm{1\over2},\pm{1\over2}}$ to fill up the ``diamond''.

\subsection{\bf Relation to semi-universal deformations}

The perturbations generated by the minus-moduli are closely related to the
concept of {\it semi-universal deformations} of singular varieties.
A {\it deformation} of a variety $K_0$ is essentially a bundle over a
certain base space $U$ with a marked point $u_0$, such that the fibre over
$u_0$ is (isomorphic to) the original variety $K_0$. The total space of
this bundle is like a space of varieties, containing the given variety
$K_0$ and others that are continuously connected to it. Precise definitions
of this concept may be found, for example, in Refs.[\REF\slodowy{P.
Slodowy, in Lecture Notes in Mathematics, 1008, ed. J. Dolgachev
(Springer-Verlag, Berlin),\hfill\break
P. Slodowy, {\it Simple Singularities and Simple Algebraic Groups},
Lecture Notes in Mathematics, 815 (Springer-Verlag, Berlin, 1980).}
\REF\tjurina{G.N. Tjurina, Math USSR Izvestiya {\bf 3} (1969)
967,\hfill\break G.N. Tjurina, Funct. Anal. Appl. {\bf 4} (1970) 68.}
\slodowy,\tjurina].

A deformation is called {\it semi-universal} if any other deformation
can be related to it by maps between the bundles and the base spaces.
In some sense, the semi-universal deformations are the ones which are
truly independent, while the others are equivalent to it by a suitable
change of variables. The application of this concept to the Kleinian
singular varieties has been worked out by Tjurina[\tjurina]. We
briefly review her results for the $A$-series, and then find a
relation with the perturbations of the ground varieties of $c=1$
string theory discussed above.

The $A_N$ Kleinian singularities are complex hypersurfaces in
\C$^3$ defined by
$$
f(X,Y,Z)\equiv Z^{2N} - XY = 0.
\eqn\ks
$$
It has been shown[\tjurina] that the semi-universal deformations of this
equation are given by the quotient $\C(X,Y,Z)/\I(f, f_X, f_Y, f_Z)$ where
$\C(X,Y,Z)$ is the ring of polynomials in $X,Y,Z$ and $\I$ is the ideal
generated by the function $f$ above, and its three partial derivatives.
This result is easy to understand to first
order in the deformation. If we consider any redefinition
$\widetilde X = X +\delta X(X,Y,Z)$, (and similarly for $Y,Z$),
then we have, to first order,
$$
f(\widetilde X, \widetilde Y, \widetilde Z) =
f(X,Y,Z) + f_X\delta X + f_Y\delta Y + f_Z\delta Z
\eqn\redef
$$
It is clear from this that such a redefinition can absorb any perturbation
containing at least one power of $f, f_X, f_Y$ or $f_Z$. Hence the
independent deformations are given by quotienting with the ideal generated
by these polynomials.
This quotient is easily computed and one finds that the semi-universal
deformation of eq.\ks\ is the $(2N-1)$-parameter family of varieties
$$
Z^{2N} - XY = t_1 Z^{2N-2} + \cdots + t_{2N-2} Z + t_{2N-1}.
\eqn\sudef
$$

Returning now to the (non-chiral) ground varieties in eq.\gvar (which
are the ones related to closed-string theory), we may ask what are the
semi-universal deformations of these.  Strictly speaking, the above
result on deformations of Kleinian varieties with isolated singular
points does not apply to the non-chiral ground varieties at the $A_N$
points, which have lines of singular points.  However, these spaces
arise as $U(1)$ quotients of products of the Kleinian varieties and
their defining equations are very similar (compare eqs.\gvar\ and
\ks), so we proceed with the assumption that here too, semi-universal
deformations are obtained by quotienting with the ideal generated by
the defining function and its first derivatives. This is anyway true
to first order in the deformation parameters, as one can simply carry
over the argument above eq.\sudef. The quotient is again
straightforward to compute, but leads this time to an
infinite-dimensional space of deformations (in accordance with the
result[\REF\matheryau{J.N. Mather and S.S.-T. Yau, Invent. Math. {\bf
69} (1982) 243.}\matheryau] that the space is finite-dimensional if
and only if the variety has an isolated singularity). The result is $$
(a_1 a_2)^N - XY = \sum_{k,l\hbox{ {\rm not both} }> (N-1)} t_{kl}
a_1^k a_2^l.
\eqn\ncsudef
$$

Now to make contact with the results of the previous section, one should
ask whether this infinite-dimensional space of semi-universal deformations
is realized by physical perturbations of the $A_N$ CFT background.
Examining eqs.\pvarplus\ and \pvarminus, we find that only a finite subset
of the perturbations in eq.\ncsudef\ above can be realised in string
theory. In fact, none of the semi-universal deformations can be generated
by plus-type moduli, while the minus-type generate the finite subset of
terms in eq.\ncsudef\ given by $k,l\le (N-1)$.

Thus we have found that {\it all perturbations generated by minus-type
moduli correspond to semi-universal deformations of the ground variety}.

Let us ask under what conditions these perturbations can smoothen the
singularity of the ground varieties. First restrict to the case of
perturbations by minus-moduli of zero matter momentum, hence
keep only the terms with $k=l$, $k,l\le N-1$, in eq.\ncsudef\ above.
Rewrite this equation as
$f(a_1, a_2, X, Y) \equiv g(a_1 a_2) - XY = 0$,
where $g$ is a polynomial function of degree $N$. The tangent space to
the perturbed variety is defined by the normal vector
$(f_{a_1}, f_{a_2}, f_X, f_Y) =
(a_2 g'(a_1 a_2), a_1 g'(a_1 a_2), -Y, -X)$.
The variety will be singular at points where this vector vanishes, and
which satisfy $f=0$. Such points fall into two
classes:
$$
\eqalign{
\hbox{\rm I :}&\qquad X=Y=0,\qquad g(a_1 a_2) =0,\qquad g'(a_1 a_2) = 0\cr
\hbox{\rm II :}&\qquad X=Y=0,\qquad g(a_1 a_2) =0,\qquad a_1 = a_2 = 0.\cr}
\eqn\conditiontwo
$$
Condition I is satisfied whenever the polynomial $g$ has a
multiple root. For each such root $r_i$, we have singularities on the
hyperbola $a_1 a_2 = r_i, X=Y=0$. Thus in this situation, the
singularities of the unperturbed variety, which lie on the pair of
straight lines $a_1 a_2 =0$[\gjm], remain present but lie on the union of
several hyperbolae, one for each multiple root. Condition II is
satisfied if the polynomial $g$ has no constant term, which means that the
perturbed variety passes through the origin. In this case, the origin is
the only singular point. If neither of the conditions
is satisfied (which is true for generic
perturbations) then the perturbed variety is nonsingular. Note that if we
perturb by the cosmological operator alone, then to first order, 0 is an
$(N-1)$-fold multiple root, so that (for $N\ge 2$) condition I is
satisfied and the singular locus remains the pair of straight lines
$a_1 a_2 = 0$. Thus, the fact that to first order, the cosmological
perturbation removes the singularity at the $SU(2)$ point[\witten] seems to
be a nongeneric case.

\subsection{\bf Calculation for perturbed ground ring action}

In this section we will calculate explicitly the perturbations of the
ground variety by the minus-moduli in some special cases. The computations
substantiate the momentum counting argument presented above.

The set of discrete plus-moduli form a faithful module of the ground
ring[\kms] and we can study the effect of perturbations
by studying the ring action on this module. Let us start with the unperturbed
theory at the $SU(2)$ point. The $a_i$ acting on a
state $Y^+_{s,n}\ybar{}^+_{s,n'}$ produce states of
appropriate momenta. Due to the orthonormality of states with
different momenta, the resulting state is uniquely determined by taking
its inner product with states in the dual module $Y^-\ybar{}^-$.
For example, $a_1Y^+_{s,n}\ybar{}^+_{s,n'}$ is a state that has
non-zero inner product only with
$Y^-_{s+{1\over2},-n-{1\over2}}\ybar{}^-_{s+{1\over2},-n'-{1\over2}}$.
We now use the $SU(2)$ relation
$Y^\pm_{s,n} = \sqrt{{(s+n)!\over(2s)!(s-n)!}}
\left(\oint J_-\right)^{s-n} Y^\pm_{s,s}$,
and deform the contour of $J_-$ to make it act on $a_i$ and $Y^-$. This
reduces our problem to knowing the ground ring action on the tachyon. One
easily finds that
$$
\eqalign{
a_1\;Y^+_{s,s}\ybar{}^+_{s,s} =
(2s)^2Y^+_{s+{1\over2},s+{1\over2}}\ybar{}^+_{s+{1\over2},s+{1\over2}}
&\qquad a_2\;Y^+_{s,s}\ybar{}^+_{s,s} =
|A(s)|^2 Y^+_{s+{1\over2},s-{1\over2}}\ybar{}^+_{s+{1\over2},s-{1\over2}}\cr
a_3\;Y^+_{s,s}\ybar{}^+_{s,s} = 2s\overline{A}(s)
Y^+_{s+{1\over2},s+{1\over2}}\ybar{}^+_{s+{1\over2},s-{1\over2}}
&\qquad a_4\;Y^+_{s,s}\ybar{}^+_{s,s} =
2s A(s) Y^+_{s+{1\over2},s-{1\over2}}\ybar{}^+_{s+{1\over2},s+{1\over2}},\cr}
\eqn\graction
$$
where $A(s)$ is a constant that we will now determine by self-consistency.
Carrying out the contour deformation in the inner product, we have
$$
a_2\;Y^+_{s,n}\ybar{}^+_{s,n'} = \sqrt{(s-n+1)(s-n'+1)}|A(s)|^2
Y^+_{s+{1\over2},n-{1\over2}}\ybar{}^+_{s+{1\over2},n'-{1\over2}}.
\eqn\atwoaction
$$
By the \Z$_2$ symmetry under $X\to -X$ of the problem, the above for
$n=n'=-s$ should be the same as the $a_1$ action on
$Y^+_{s,s}\ybar{}^+_{s,s}$. This determines
$A(s) = {2s\over \sqrt{2s+1}}$.
(We have chosen the positive square root so as to make the
ring action on the module commutative.) This finally gives us the ground
ring action on discrete states:
$$\eqalign{
a_{1(2)}\;Y^+_{s,n}\ybar{}^+_{s,n'} &= {(2s)^2\over 2s+1}
\sqrt{(s\pm n+1)(s\pm n'+1)}
Y^+_{s+{1\over2},n\pm{1\over2}}\ybar{}^+_{s+{1\over2},n'\pm{1\over2}},\cr
a_{3(4)}\;Y^+_{s,n}\ybar{}^+_{s,n'} &= {(2s)^2\over 2s+1}
\sqrt{(s\pm n+1)(s\mp n'+1)}
Y^+_{s+{1\over2},n\pm{1\over2}}\ybar{}^+_{s+{1\over2},n'\mp{1\over2}},\cr
}
\eqn\gractionstates
$$
{}From eqs.\gractionstates, it is easy to check the commutativity
of the ring and the relation \gcone. Note also that the above
equations agree with the explicit calculations of Ref.\REF\wuzhu{Y.-S.
Wu and C.-J. Zhu, Preprint UU-HEP-92-6, hep-th/9209011.}[\wuzhu]
after setting the new moduli to zero there.

Now consider the perturbation by the cosmological operator. The inner
product that we are considering receives corrections due to the presence of
the (integrated) cosmological operator. The corrections to the $a_1$, $a_3$
and $a_4$ actions on the discrete tachyon $Y^+_{s,s}\ybar{}^+_{s,s}$ vanish
by simple momentum counting --- they would have produced states whose
$n$-value is greater than the $s$-value, and such a state is not in the
cohomology. Only the $a_2$ action is non-zero:
$$
a_2\;Y^+_{s,s}\ybar{}^+_{s,s}\Big|_{correction}
= - \mu a_2 Y^+_{s,s}\ybar{}^+_{s,s}\int W_{0,0,0}
= \mu B(s)Y^+_{s-{1\over2},s-{1\over2}}\ybar{}^+_{s-{1\over2},s-{1\over2}},
\eqn\atwocorr
$$
where $B(s)$ is a constant to be determined. we repeat the steps as in the
unperturbed case, and fix $B(s)$ by demanding that the perturbed ring is
commutative. This determines $B(s)$ uniquely to be
$B(s) = 2s/(2s+1)(2s-1)^2$.
Hence the corrections to the ground ring action are:
$$\eqalign{
a_{1(2)}\;Y^+_{s,n}\ybar{}^+_{s,n'}\Big|_{correction}
&= \mu {\sqrt{(s\mp n)(s\mp n')}\over (2s+1)(2s-1)^2}\;
Y^+_{s-{1\over2},n\pm {1\over2}}\ybar{}^+_{s-{1\over2},n'\pm{1\over2}},\cr
a_{3(4)}\;Y^+_{s,n}\ybar{}^+_{s,n'}\Big|_{correction}
&= \mu {\sqrt{(s\mp n)(s\pm n')}\over (2s+1)(2s-1)^2}\;
Y^+_{s-{1\over2},n\pm {1\over2}}\ybar{}^+_{s-{1\over2},n'\mp{1\over2}},\cr
}
\eqn\correctgraction
$$
Combining eqs.\gractionstates\ and \correctgraction, we get
$(a_1a_2-a_3a_4)Y^+_{s,n}\ybar{}^+_{s,n'} = \mu Y^+_{s,n}\ybar{}^+_{s,n'}$,
verifying the cosmologically perturbed ring relation \pcone\ at
the $SU(2)$ point.

It does not appear to be straightforward to extend these calculations to
the other $A_N$ points, although that would provide a useful check on our
conclusions. We can, however, study the action of the ground ring
elements on the intermediate tachyons --- ($N-1$ species of these are
present in the $A_N$ theory for $N\ge 2$) --- and corrections to them.
Since these form an unfaithful module (for each species of tachyon)[\kms],
there could be additional relations in the module that are not ring
relations. Indeed, the fermi surface equation $a_1 a_2=0$ is true only for
the intermediate tachyons and not a relation in the (unperturbed) ground
ring.

Again, start by considering the effect of the cosmological
perturbation. Under this
perturbation, the action of the ground ring generators on the tachyon
receives corrections of the form $-\mu a_i \tet_q\int W_{0,0,0}$. A simple
momemtum counting shows once again that the corrections to $a_1$, $a_3$
and $a_4$ on the tachyon vanish --- the would-be state is not in the
cohomology. The correction to $a_2$ is the tachyon with momentum $q-1$.

Following[\kachru],
this correction can be written as:
$$
\eqalign{ -\mu\lim_{w\to 0}a_2(w,\bar w)\tet_q(0)\int d^2z\;
W_{0,0,0}(z,\bar z) &= -\mu\lim_{w\to 0}\int d^2z
|w|^{2-2q}|z|^{2q-4}\tet_{q-1}(0)\cr &=
-\mu\pi{\D(q-1)\over\D(0)\D(q)}\tet_{q-1}(0),\cr}
\eqn\mupert
$$
where $\D(x) = \G(x)/\G(1-x)$.
To simplify the above formula we define the following (singular)
normalization of the tachyons and the couplings $u^-$ corresponding to the
perturbation by the minus-moduli
$$
\eqalign{
    \widetilde{\tet}_q &= \D(q)\tet_q\cr
    \tilde{u}^-_{s,n} &= {\pi\over\D(-2s)} u^-_{s,n}.\cr}
\eqn\renorm
$$

With these normalizations, the perturbed action of $a_2$ on the
tachyon is $a_2\;\widetilde{\tet}_q = -\tilde{\mu}\widetilde{\tet}_{q-1}$.
On the other hand, action of $a_1$ on the normalized tachyon is
$a_1\widetilde{\tet}_q = -\widetilde{\tet}_{q+1}$.
Combining these, we find
$$
a_1 a_2 \widetilde{\tet}_q
=\tilde{\mu}\;\widetilde{\tet}_q
\eqn\modulecone
$$
the equation for the perturbed fermi surface[\kachru,\barbon].

We can now study the effect of perturbations corresponding to other
minus moduli. Consider the modulus $W^-_{1,0,0}$, which
corresponds to the black hole perturbation[\blackholepert,\mmsbh] and
produces a constant term in the equation of the perturbed $A_2$
variety\pvarminus. Momentum-counting shows in particular that the
correction to the action of $a_2$ on the tachyon,
$-u^-_{1,0}a_2\tet_q\int W^-_{1,0,0}$, would result in a state that is
not in the cohomology --- which implies that this correction vanishes.
Despite this, the correction to the action of $a_2^2$ on this tachyon
is expected to be nonzero by our arguments. This is not really a
contradiction, since we are considering the products of various fields
in the cohomology when their locations on the worldsheet collide, and
the decoupling of BRS-trivial fields can fail to hold at coincident
points.

Using the explicit expression for $a_2^2 = {\cal O}_{1,-1}
\overline{{\cal O}}_{1,-1}$, a straightforward though somewhat tedious
calculation gives
$$
\eqalign{
&
-u^-_{1,0}\lim_{w\to 0}a_2^2(w,\bar w)\tet_q(0)
\int d^2z\;W^-_{1,0,0}(z,\bar z)
\cr
&\qquad\qquad\qquad= -u^-_{1,0}\int d^2z |z|^{2q-4}|1-z|^2
\left|\left({q/\sqrt 2\over z} + {3/\sqrt2\over (1-z)}\right)\right|^2\;
\tet_{q-2}(0)\cr&\qquad\qquad\qquad
= u^-_{1,0}{\pi\D(-2)^{-1}\over 8(q-1)^2(q-2)^2}
\tet_{q-2}(0),
\cr}
$$
which after normalization \renorm\ gives the correction due to the black
hole perturbation
$$
a_2^2\widetilde{\tet}_q = {1\over 8}\tilde{u}^-_{1,0}
\widetilde{\tet}_{q-2},
\eqn\bholepert
$$
which gives $a_1^2a_2^2\tet_q={1\over8}\tilde{u}^-_{1,0}\tet_q$.

\subsection{\bf Discussion and Conclusions}

We have studied the deformation of the $A_N$ ground varieties of the $c=1$
string theory under perturbations by the physical moduli of the theory
using a simple scaling relation. The unperturbed ground varieties were
related to the the Kleinian singular varieties[\gjm]. Here we find that a
finite set of physical perturbations produce the the analogue of the
semi-universal deformation of the corresponding ground variety.

Let us briefly comment on the question of integrability of the
perturbations that we have been considering.  For the plus-type
perturbations, the requirement of integrability implies that eq.\pvarplus\
should be taken seriously only for $n=0$, in which case the right hand
side is a power series in $(a_1 a_2)$ with powers ranging from $N-1$ to
infinity. The first term is associated to the cosmological perturbation.
The second one, proportional to $(a_1 a_2)^N$, is associated to the
operator $W^+_{1,0,0}= \del X \bar\del X$ which is the radius-changing
perturbation.  This term can clearly be absorbed into the unperturbed
equation by a simple rescaling of $a_1$ and $a_2$. Thus the ground variety
is unaffected, to first order, by a radius-changing perturbation. This may
appear a little disturbing at first, since the ground variety of the
theory clearly depends on the compactification radius. However, the
radius-dependence of this variety, which has been analysed in
Ref.[\gjm], is highly singular.  Even the dimension of the variety
depends on whether the radius is rational or irrational. This phenomenon
clearly cannot be perturbative in the radius, which explains why we do not
see it here. Naturally, this demonstrable limitation of perturbation
theory casts some doubt on the validity of any analysis which is
perturbative in the $u^\pm$.

As for the minus-type perturbations, the requirement of integrability
presumably does not impose any limitation on them, since their chiral
ingredients $W^-_{s,n}$ satisfy an abelian algebra. It has in fact
been shown explicitly in Ref.[\mmsbh] that the minus-moduli of zero
matter momentum generate all-orders solutions to the string field
theory equations of motion, which amounts to a proof of their
integrability. Thus eq.\pvarminus\ is meaningful for all $s,n$ in the
``diamond'' of Fig.1.

The special role played by moduli of minus-type, which include the
physically interesting black hole perturbation, seems worth
understanding in more detail. Recent attempts to find the black hole
perturbation in the framework of matrix
models\REF\matbh{S.~R. Das, Mod. Phys. Lett. {\bf A8}
(1993) 69; {\bf A8} (1993) 1331,\hfill\break
A. Dhar, G. Mandal and S. Wadia, Mod. Phys. Lett. {\bf A7}
(1992) 3703; Preprint TIFR/TH/93-05, hep-th/9304072, \hfill\break
J.~G. Russo, Phys. Lett. {\bf B300} (1993) 336, \hfill\break
Z. Yang, Preprint UR-1251, hep-th/9202078,\hfill\break
A. Jevicki and T. Yoneya, Preprint
NSF-ITP-93-67, BROWN-HEP-904,\hfill\break
UT-KOMABA/93-10, hep-th/9305109.}[\matbh]
may help to illuminate this question.

\noindent {\bf Acknowledgements:} We would like to thank D.P. Jatkar for
collaboration at an early stage of this work. It is a pleasure to thank A.
Parameshwaran, K. Paranjape and A. Sen for valuable discussions.

\refout
\bye